\documentclass[aps,prl,twocolumn,groupedaddress]{revtex4-1}
\usepackage{amsmath,graphicx}

\def\be{\begin{equation}}
\def\ee{\end{equation}}
\def\bea{\begin{eqnarray}}
\def\eea{\end{eqnarray}}
\def\eps{\epsilon}

\def\eps{\epsilon}

\def\cO{  {\cal O}  }

\begin{document}

\title{Multiloop integrals in dimensional regularization made simple}

\author{Johannes M.\ Henn}
\email[]{jmhenn@ias.edu}
\affiliation{ Institute for Advanced Study, Princeton, NJ 08540, USA}


\begin{abstract}
Scattering amplitudes at loop level can be expressed in terms of Feynman integrals. 
The latter satisfy partial differential equations in the kinematical variables.
We argue that a good choice of basis for (multi-)loop integrals can lead to significant simplifications of the differential equations, and propose criteria for finding an optimal basis. This builds on experience obtained in supersymmetric field theories that can be applied successfully to generic quantum field theory integrals. It involves studying leading singularities and explicit integral representations. When the differential equations are cast into canonical form, their solution becomes elementary. The class of functions involved is easily identified, and the solution can be written down to any desired order in $\eps$ within dimensional regularization. Results obtained in this way are particularly simple and compact. In this letter, we outline the general ideas of the method and apply them to a two-loop example. 
\end{abstract}

\pacs{12.38.-t,03.70.+k, 11.10.-z}

\maketitle

\section{Introduction and summary}
\label{sec-intro}

Scattering amplitudes are of fundamental interest in quantum field theory,
as they connect theory to (collider) experiments.
A theoretical challenge is to describe events with many jets, as well as to
increase the precision of the predictions by going beyond leading and next-to-leading 
order accuracy, see \cite{Blumlein:2012gu} for a recent discussion.

Computing scattering amplitudes usually consists of two steps.
The first is to find an expression in terms of (Feynman) loop integrals, 
and the second is to evaluate the latter.
In supersymmetric theories, the first step is essentially solved by virtue of 
(generalized) unitarity or recursion relations \cite{Bern:1994zx,Britto:2005fq,ArkaniHamed:2010kv},
where the idea is to reconstruct the loop integrand from its analytic 
behavior, in particular factorization on propagator poles. 
In developing these ideas, the maximally supersymmetric Yang-Mills theory was instrumental. 
What can we learn from recent advances in this theory at loop level?

Although many impressive perturbative results are available,
unfortunately  the methods used to obtain them 
cannot be applied directly to generic scattering amplitudes. 
The reason is that most of them relied in some aspect on the duality 
between scattering amplitudes and Wilson loops specific to that theory.
Nonetheless, these advances suggest to us that there should also be 
easier ways to directly carry out the Feynman integrations. 
Progress in this direction commenced with a better understanding of loop integrands.
It was observed that new representations for the latter \cite{ArkaniHamed:2010kv,ArkaniHamed:2010gh} lead to simple analytic answers \cite{Drummond:2010mb,Dixon:2011nj},
as a consequence of differential equations they satisfy \cite{Drummond:2010cz}.
Moreover, mathematics for iterated integrals \cite{arXiv:math/0606419,2009arXiv0908.2238G,Goncharov:2010jf} furthered the understanding of the transcendental functions involved. 

In this letter, we apply these ideas to the evaluation of generic
loop integrals in dimensional regularization, and combine them with state-of-the art
techniques of QCD. 
Our method applies to planar or non-planar, massless or massive integrals equally.

As we review shortly, the calculation of an arbitrary loop integral can always be related to the calculation
of a finite set of master integrals. The question we wish to address here is how to choose a `good' set of master integrals.

One key idea is that we would like loop integrals to have simple properties
under the action of differential operators.
In order to define what is meant by simple, let us introduce the
concept of {\it degree of transcendentality} $\mathcal{T}(f)$ of a function $f$, which applies
to a large class of iterated integrals \cite{arXiv:math/0606419,2009arXiv0908.2238G}.
$\mathcal{T}(f)$ is defined as the number of iterated integrals needed to define
the function $f$, e.g. $\mathcal{T}(\log)= 1$, $\mathcal{T}({\rm Li}_{n}) = n$, etc.
We also have $\mathcal{T}(f_1 f_2) = \mathcal{T}(f_1) +  \mathcal{T}(f_2)$.
Constants obtained at special values are also assigned transcendentality, e.g. $\mathcal{T}({\zeta}_{n}) = n$.
Algebraic factors have degree zero.
Functions that we will be interested in have {\it uniform} (degree of) transcendentality, i.e.
if $f$ is a sum of terms, all summands have the same degree.

Moreover, we call such functions {\it pure} if their degree of transcendentality is lowered by taking a derivative, i.e. $\mathcal{T}(d \, f) = \mathcal{T}(f) - 1$. This implies that the transcendental functions in $f$ cannot be multiplied by algebraic coefficients, which would otherwise be `seen' by the differential operators.

There are several guiding principles that can help to find such integrals. 
Although to the best of our knowledge there is no general proof,
it has been observed that integrals having constant {\it leading singularities} 
\cite{Cachazo:2008vp,ArkaniHamed:2010gh} have these properties. 
The latter are defined by analytically 
continuing the momenta to complex values, and replacing the integration
over space-time by contour integrals around poles of the integrand.
Another way the properties discussed above 
can be made manifest is when an appropriate `d-log' representation
is available \cite{ArkaniHamed:2012nw}, where the integrand is written as a 
logarithmic differential form. 
This approach works particularly well for Wilson loop integrals.
Finally, we have also found explicit representations 
based on Feynman parameters to be useful.

One might think that few Feynman integrals have such nice properties. 
In fact, as we argue here, quite the opposite is the case.
This claim is supported by many examples in the literature. As a specific case in point, the 
above ideas were employed in \cite{Gehrmann:2011xn} to present 
massless planar and non-planar form factor integrals in a basis
where each integral has uniform transcendentality.

Having discussed these general ideas, let us return to the issue of the
reduction to master integrals and the calculation of the latter.

Feynman integrals can be classified according to their topology,
starting with the integrals where the maximal number of propagators is present.
The propagators, labelled by $i$, are raised to powers $a_{i}$.
Subtopologies, where certain propagators are absent, are obtained by setting 
the corresponding indices $a_{i}$ to zero.

For each topology, there is a set of integration-by-parts (IBP) identities \cite{Chetyrkin:1981qh} that relates integrals with different values of the $a_{i}$. 
These equations follow from the Poincar{\'e} invariance of the integrals, 
which is preserved in dimensional regularization.
They are linear in the integrals, with the coefficients being
rational functions of the kinematical invariants and the space-time dimension.
In this way, one can relate an integral with general integer powers to a finite set of master, or basis integrals.
In practice, a set of basis integrals can be found straightforwardly by using various public computer codes \cite{Anastasiou:2004vj,*Smirnov:2008iw,*Smirnov:2013dia,*Studerus:2009ye,*vonManteuffel:2012np}. For a recent review and more details and references, see \cite{Smirnov:2012gma}.

Having reduced the general problem in this way, 
one would then like to compute the basis integrals.
From the discussion above it should not be surprising that 
we will use the method of differential 
equations \cite{Kotikov:1990kg,Kotikov:1991pm,Bern:1993kr,Gehrmann:1999as}. 
The idea is to differentiate w.r.t. the kinematical invariants.
This can be implemented on the Feynman integrals by defining appropriate derivatives
w.r.t. the momenta (respecting momentum conservation and on-shell conditions.)
The r.h.s. of such an equation involves integrals of the same topology, but with different powers $a_{i}$. The latter integrals can then be re-expressed via the IBP relations in terms of the chosen basis integrals.
So in general we obtain a set of linear, first-order partial differential equations for the basis integrals.

Denoting the kinematical variables by $x_{i}$, the set of $N$ basis integrals by $f_{i}$, and working in $D=4-2\eps$ dimensions, this set of equations takes the form
\begin{align}\label{diffeq_general}
 \partial_m
 f(\eps, x_{n} ) = A_{m}(\eps, x_{n}) f(\eps, x_{n}) \,,
\end{align}
where $\partial_{m} = \frac{\partial}{\partial x_m }$, and
each $A_{m}$ is an $N \times N$ matrix.
The matrices have to satisfy the integrability conditions (except possibly for special singular values of the $x_{m}$),
\begin{align}\label{integrability_general}
\partial_{n} A_{m} - \partial_{m} A_{n}  + [ A_{n} ,A_{m} ] =0\,, 
\end{align}
where $[A,B]:=AB-BA$.

In order to completely specify the solution, one has to provide a boundary condition. This can in general be done by studying physical limits of the scattering process under consideration.

In practice, one would like to solve eq. (\ref{diffeq_general}) in 
a Laurent expansion around $\eps = 0$ \footnote{In some cases, one can find a solution exact in $\eps$ in terms of hypergeometric functions and their generalizations. However, then one faces the complicated task of deriving the $\eps$ expansion of those functions.}.

Under a change of basis $f \to B f$, the matrices $A_{m}$ transform as
\begin{align}\label{eq_changevar}
A_{m} \longrightarrow B^{-1}A_{m}  B - B^{-1} (\partial_{m}  B) \,.
\end{align}
Note that here $B$ can be in principle any $N\times N$ matrix, where each entry is a function of $\eps$ and of the kinematical variables $x_{n}$. 

Equation (\ref{diffeq_general}) can simplify considerably when a good choice of basis is made.
Our conjecture is that an optimal choice of integral basis $f_{i}$ can be reached, 
where the integration of the system of differential equations becomes trivial,
in the sense explained below.

One could imagine several simplified versions of eq. (\ref{diffeq_general}). 
We studied various cases of practical interest, 
and based on that evidence we propose 
that one can transform eq. (\ref{diffeq_general}) 
into the following form,
\begin{align}\label{diffeq_special}
\partial_{m} f(\eps, x_{n} ) = \eps \, A_{m}(x_{n}) f(\eps, x_{n} ) \,.
\end{align}
We remark that in this case, the integrality condition (\ref{integrability_general}) becomes
\begin{align}\label{integrability_special}
\partial_{n} A_{m} - \partial_{m} A_{n} = 0   \,, \qquad [ A_{n} ,A_{m} ] =0\,.
\end{align}
For the discussion of the properties of the solution
it is convenient to combine eqs. (\ref{diffeq_special}) and write them in differential form,
\begin{align}\label{diffeq_special2}
d \, f(\eps, x_{n} ) = \eps \,  d \, \tilde{A}(x_{n}) \,  f(\eps, x_{n} )
\end{align}
We may also note that one can formally solve eq. (\ref{diffeq_special2}) in terms of a path-ordered exponential, 
\begin{align}
f = P e^{\eps \int_{\mathcal{C}} d \tilde{A} } f(\eps =0)\,,
\end{align} 
where the integration contour $\mathcal{C}$ connects the
base point (representing the boundary condition) to $x_n$.
In other words, the perturbative solution in $\eps$ is given by iterated integrals,
where the entries of $d \tilde{A}$ determine the integration kernels.

We stress that once a form (\ref{diffeq_special}), or equivalently (\ref{diffeq_special2}) is reached, then the problem of solving for the
master integrals $f_{i}$ in the $\eps$ expansion is essentially solved.

The form (\ref{diffeq_special2}) of the equations can also make the transcendentality
properties of the solution manifest. 
In order to see this, let us introduce one new concept. In dimensional regularization,
it is customary to assign degree of transcendentality $-1$ to $\eps$.
In this way, one can discuss the transcendentality properties of functions appearing 
in the Laurent expansion of $\eps$-dependent expressions. For example, in
$x^{\eps} = 1 + \eps \log x  + \cO(\eps^2)$
each summand has the same degree of transcendentality, namely zero.
Then, if $d \,\tilde{A}$ in (\ref{diffeq_special2}) is a logarithmic one-form, it 
is clear that the answer will have uniform transcendentality, to all perturbative orders in $\eps$.

Along the same lines, one can immediately determine from eq. (\ref{diffeq_special2}) which class
of functions the solution will be expressed in. This is best discussed using the notion of {\it symbol}
of a transcendental function \cite{arXiv:math/0606419,2009arXiv0908.2238G,Goncharov:2010jf}.
The entries of $\tilde{A}$ in eq. (\ref{diffeq_special2}) determine the alphabet of the symbols of the solution, again to all orders in $\eps$.  

In principle, starting with a random basis of master integrals $f$, one could attempt to
find an appropriate set of functions $B$ in eq. (\ref{eq_changevar}) in order to reach the canonical form (\ref{diffeq_special}).
However, this seems like a formidable task in general, 
and for that reason the criteria outlined above are very useful in practice.

In the remainder of this letter, we present a nontrivial example of this method.
A more detailed discussion, as well as further applications, will be given elsewhere.

\subsection{Planar two-loop master integrals for $2\to 2$ scattering}
We consider the planar double ladder integrals \cite{Smirnov:1999gc,Gehrmann:1999as}.
One can see via IBP that these constitute all loop integrals for virtual corrections to
massless $2 \to2$ scattering, in any gauge theory.
We introduce the notation
\begin{align}\label{double ladder}
I_{a_{1}, \ldots ,a_{9}} := e^{2 \eps \gamma_{\rm E}} \int \frac{d^{D}k_{1}d^{D}k_{2}}{ (i\pi^{D/2})^2} 
\prod_{m=1}^{9} (P(q_m))^{a_{m}}\,,
\end{align}
with the propagator $P(q):= 1/(-q^2)$, and the set of possible momenta $q_{m}$,
corresponding to $m=1,\ldots 9$, respectively, is
$k_1, k_1+p_1, k_1+p_{12}, k_1+p_{123} , k_2, k_2+p_{12} ,k_2+p_{123},k_1-k_2$,
where $p_{12}=p_1+p_2$ and $p_{123}=p_{1}+p_{2}+p_{3}$.
We have $p_{i}^2=0$ and 
$\sum_{i=1}^{4} p_{i}=0$. The results depend on the Mandelstam variables
$s=2 p_1\cdot p_2$ and $t=2 p_2 \cdot p_3$.

\begin{figure}[ht]
\includegraphics{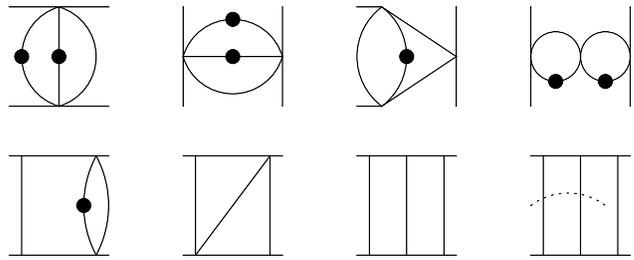}%
\caption{Integral basis corresponding to $f_{1},\ldots f_{4}$ (first line) and $f_{5}, \ldots f_{8}$ (second line), up to overall factors. Fat dots indicate doubled propagators, and the dotted line an inverse propagator. The incoming momenta are labelled in a clockwise order, starting with $p_{1}$ in the lower left corner.
\label{fig:4ptints}}
\end{figure}

There are $8$ master integrals for this problem. 
We choose the following basis, see Fig.~\ref{fig:4ptints},
\begin{align}
\allowdisplaybreaks
f_1 &= - \eps^2 \, (-s)^{2 \eps} \, t\, I_{0,2,0,0,0,0,0,1,2} \,, \\
 f_2 &=\eps^2 \, (-s)^{1+2 \eps} \,   I_{0,0,2,0,1,0,0,0,2}  \,, \\
f_3 &= \eps^3 \, (-s)^{1+2 \eps}   \, I_{0,1,0,0,1,0,1,0,2}\,, \\
 f_4 &=- \eps^2 \, (-s)^{2+2 \eps} \,   \,  I_{2,0,1,0,2,0,1,0,0}\,,  \\
f_5 &= \eps^3 \, (-s)^{1+2 \eps} \,  t \, I_{1,1,1,0,0,0,0,1,2}\,, \\
 f_6 &=- \eps^4 \, (-s)^{2 \eps} \,   (s+t) \, I_{0,1,1,0,1,0,0,1,1}\,, \\
f_7 &= - \eps^4 \, (-s)^{2+2 \eps} \,  t\, I_{1,1,1,0,1,0,1,1,1} \,, \\
 f_8 &=- \eps^4 \, (-s)^{2+2 \eps}  \, I_{1,1,1,0,1,-1,1,1,1}  \,.
\end{align}
Here the normalization was chosen such that all $f_i$ admit a Taylor expansion in $\eps$. 
Moreover, they depend on $s$ and $t$ through
the dimensionless variable $x=t/s$ only.
For this choice of basis, we find
\begin{align}\label{derivativeint}
\partial_{x} f = \eps \left[ \frac{a}{x} + \frac{b}{1+x}  \right]\, f \,, 
\end{align}
with the constant matrices
\begin{align}
a &= \left(
\begin{array}{cccccccc}
 -2 & 0 & 0 & 0 & 0 & 0 & 0 & 0 \\
 0 & 0 & 0 & 0 & 0 & 0 & 0 & 0 \\
 0 & 0 & 0 & 0 & 0 & 0 & 0 & 0 \\
 0 & 0 & 0 & 0 & 0 & 0 & 0 & 0 \\
 \frac{3}{2} & 0 & 0 & 0 & -2 & 0 & 0 & 0 \\
 -\frac{1}{2} & \frac{1}{2} & 0 & 0 & 0 & -2 & 0 & 0 \\
 -3 & -3 & 0 & 0 & 4 & 12 & -2 & 0 \\
 \frac{9}{2} & 3 & -3 & -1 & -4 & -18 & 1 & 1
\end{array}
\right)\,, 
\end{align}
and
\begin{align}
b &= \left(
\begin{array}{cccccccc}
 0 & 0 & 0 & 0 & 0 & 0 & 0 & 0 \\
 0 & 0 & 0 & 0 & 0 & 0 & 0 & 0 \\
 0 & 0 & 0 & 0 & 0 & 0 & 0 & 0 \\
 0 & 0 & 0 & 0 & 0 & 0 & 0 & 0 \\
 -\frac{3}{2} & 0 & 3 & 0 & 1 & 0 & 0 & 0 \\
 0 & 0 & 0 & 0 & 0 & 2 & 0 & 0 \\
 3 & 6 & 6 & 2 & -4 & -12 & 2 & 2 \\
 -\frac{9}{2} & -3 & 3 & -1 & 4 & 18 & -1 & -1
\end{array}
\right)\,.
\end{align}
Equation (\ref{derivativeint}) is a simple instance of the Knizhnik-Zamolodchikov equations \cite{Knizhnik:1984nr}\footnote{We remark that the same equations also appeared recently in the context of the 
$\alpha'$ expansion of tree-level superstring amplitudes, see arXiv:1304.7304 [hep-th].}.
The three singular points $\{ 0,-1,\infty \}$ for $x$ correspond to the physical limits $s=0, u=-s-t=0$ and $t=0$, respectively. 

Taking into account that the leading term in the $\eps$ expansion must be a constant,
it follows from eq. (\ref{derivativeint}) 
that the result at any order
in $\eps$ can be written as a linear combination of harmonic polylogarithms \cite{Remiddi:1999ew} of argument $x$, 
with indices drawn from the set $\{0,-1\}$. In particular, the symbol alphabet in this case is $\{ x, 1+x\}$.
Finally, the requirement that the
planar integrals be finite at $x=-1$ and real-valued for $x$ positive turns out to fix all except
two boundary constants. The latter can be related to the trivial propagator-type
integrals $f_2$ and $f_4$, which are known in closed form.
This completely solves this family of Feynman integrals, to all orders in $\eps$.
We see that all basis elements $f_{i}$ have uniform degree of transcendentality, to all orders in $\eps$.
One may verify agreement with ref. \cite{Smirnov:1999gc}, to the order in $\eps$ computed there.

\section{Discussion and outlook}

In this letter, we argued that a good basis choice for master integrals can
significantly simplify the differential equations they satisfy.
We motivated and discussed guiding principles for choosing good master integrals, based on
their transcendentality properties. The latter can be discussed through their leading
singularities. We remark that the basis choice is not unique.

We provided a non-trivial example at the two-loop level \cite{Smirnov:1999gc,Gehrmann:1999as}. These integrals can now be trivially obtained to any order in $\eps$. 
All master integrals have uniform degree of transcendentality and are given by compact expressions.

It would be interesting to find criteria for, or prove or disprove 
the existence of a matrix $B$ of 
eq. (\ref{eq_changevar}) that leads to (\ref{diffeq_special}).
We would like to stress that beyond the example given here, we found this method to apply
to many further cases of practical interest.
In particular, we expect applications to previously unknown integrals 
involving top quarks or Higgs particles, or to
Bhabha scattering, to name a  few examples. 
Preliminary results also show that the method can be applied successfully
to integrals in heavy quark effective theory.

In more complicated multi-leg processes, or processes involving masses, the appropriate set of integral functions may not
yet be known. We anticipate that our method will be a convenient way of solving this problem,
and lead to investigations of generalized functions appropriate for those scattering processes.
In this context we also wish to stress that the differential equations can first be trivially
solved in terms of symbols, and possible simplifications identified, before the problem
of finding a convenient integral representation is addressed.

A further promising avenue of research is the systematic investigation of leading singularities
in $D=D_{0}- 2 \eps$ dimensions, where $D_{0}$ is some integer. In this context, we would also like to point out that we found propagators with doubled or higher powers useful in choosing master integrals.
We focused on expansions near four dimensions, but one can apply our method in other dimensions as well, where a different choice of basis may be appropriate.

\section{Acknowledgements}
We thank S.~Caron-Huot and D.~Skinner for useful conversations,
T.~Gehrmann and N.~Arkani-Hamed for discussions and comments on the draft, 
and G.~Korchemsky and P.~Marquard for discussions and work on related topics. 
JMH was supported in part by the Department of Energy grant DE-FG02-90ER40542,
and by the IAS AMIAS fund.

\bibliography{qcdintegrals}

\end{document}